# Numerical simulation of atmospheric transport and dispersion of *Phakopsora pachyrhizi* urediniospores in South America using the state of Paraná-Brazil as a model

Eduardo Oliveira Belinelli[a], Lucas Henrique Fantin[b], Marcelo Giovanetti Canteri[b], Karla Braga[b], Eliandro Rodrigues Cirilo[a], Neyva Maria Lopes Romeiro[a], Paulo Laerte Natti[a*]

[a] Mathematics Department, Universidade Estadual de Londrina (UEL), Rodovia Celso Garcia Cid, Londrina, Paraná, Brazil

[b] Agronomy Department, Universidade Estadual de Londrina (UEL), Rodovia Celso Garcia Cid, Londrina, Paraná, Brazil

*Corresponding author

E-mail address: plnatti@uel.br

Eduardo Oliveira Belinelli   https://orcid.org/0000-0002-5925-086X

Lucas Henrique Fantin   https://orcid.org/0000-0002-5632-3007

Marcelo Giovanetti Canteri   https://orcid.org/0000-0002-6625-5909

Karla Braga   https://orcid.org/0000-0002-1794-8266

Eliandro Rodrigues Cirilo   https://orcid.org/0000-0001-7530-1770

Neyva Maria Lopes Romeiro   https://orcid.org/0000-0002-3249-3490

Paulo Laerte Natti   https://orcid.org/0000-0002-5988-2621

**Abstract**

*Phakopsora pachyrhizi* is a biotrophic fungus that needs living plant tissue in order to survive for periods in the wild. The fungus causes Asian rust and costs billion US dollars every year for its control in South American soy production. Despite the regulatory measure that prohibits the cultivation of soybeans in some months of the year (soybean-free period) in Brazil, the presence of soybean production areas in neighboring countries, such as Paraguay and Bolivia, can help the survival of the pathogen between crop seasons. It is known that *P. pachyrhizi* urediniospores can be spread/transported thousands of kilometres by the wind. In this context, the objective of this work was to develop a mathematical model to simulate the atmospheric transport of *P. pachyrhizi* urediniospores from the west, Paraguay/Bolivia to Paraná/Brazil, through storms coming from cold fronts. The transport of urediniospores was modeled by a diffusive-convective-reactive equation. The mathematical model was discretized by the finite difference method. The algebraic system resulting from discretization was solved by the Gauss-Seidel iterative method. Wind direction and the velocity of cold fronts that crossed Paraná state between October 2018 and February 2019 were used. For validation, real cases of rust occurrence in the Paraná state informed by the Anti-rust Consortium Portal in the season 2018/19 were used. A total of nine cold fronts occurred in the studied period. The wind direction varied between months. The results confirm mathematically that it is possible for urediniospores from infected areas located in a country on the west to be transported and deposited on the east, in the state of Paraná. The first case of soybean rust in Paraná state/Brazil was registered 10 days after the first cold front, suggesting that the transported and deposited urediniospores were still viable for host infection. This work reinforces the importance of the establishment of the soybean-free period in other soybean produced countries. It will also provide a better understanding of the fungus dispersion system, potentially enabling the correct use of fungicides.

**1. Introduction**

The fungus *Phakopsora pachyrhizi* Sid. & P. Sid., causal agent of Asian rust is the most important plant pathogen in Brazil (Godoy et al., 2016; Mello et al., 2021). In Brazil, every season the fungus costs approximately 2 billion US dollars due to fungicide applications (Godoy et al., 2016). Meta-analytic estimates indicated yield reduction of 60 kg.ha$^{-1}$ for 3 % of disease severity (Dalla Lana et al., 2015), and without measurement control disease losses can reached 90%. The fungus is an obligate parasite. The main hosts of *Phakopsora pachyrhizi* fungus are plants of *Fabaceae* family, such as soybean (*Glycine max* (L.) Merrill), dry bean (*Phaseolus vulgaris* L.) and kudzu (*Pueraria lobata* (Willd.) Ohwi) (Harmon et al., 2005; Lynch et al., 2006; Schneider et. al., 2005).

In this context, the soybean-free period proposed in 2005 was a regulatory programme stablished by Brazilian Ministry of Agriculture, Livestock and Food Supply (MAPA) to eliminate the main host and reduce the inoculum (Godoy et al., 2016). Despite regulatory measures, Asian rust epidemic cases have been occurring systematically in all seasons, suggesting the maintenance of host/inoculum in the off-season (Minchio et al., 2016, 2018). Multiple reasons can be related to the failure of the regulatory programme, such as the maintenance of inoculum in involuntary plants or the non-adoption of soybean-free period in some Brazilian states and soybean producing neighboring countries, such as Paraguay and Bolivia. In this context the main question is: where is the main source of primary inoculum of *P. pachyrhizi* responsible for the Asian rust epidemic in Paraná state?

The primary inoculum is a key for the development of disease epidemics and the knowledge of movement of pathogens in atmosphere is important to develop management control strategies (Aylor, 1990). It is known that fungal urediniospores can be spread thousands of kilometres by the wind, remaining viable for infection (Aylor, 2003; Isard et al., 2005; Mims and Mims, 2004; Nagarajan and Singh, 1990; Prussin et al., 2015; Schmale and Ross, 2015). Because of the soybean-free period, farmers are only allowed to plant soybean after September 15$^{th}$ in Paraná state. In this period, the dominant atmospheric current in Paraná is formed by the Southeast Trade Winds, displacements of masses of hot and humid air that move towards the areas of lower atmospheric pressure in the equatorial zone, with winds from east to west (Kalnay et al., 1996; Reboita et al., 2018). However, during the cold fronts which are responsible for rain, the air moves from west to east, which in theory would allow urediniospores transport from west countries to Paraná State/Brazil. It should be noted that the speed of Trade Winds is around 10 km/h, while wind speeds from cold fronts reach up to 90 km/h

A numerical simulation approach was used to demonstrate the potential for long-distance transport of spores, as in the work already developed by Aylor (2003), Isard et al. (2005, 2011), and Wen et al. (2017). In this work, a partial differential equation (PDE) was adopted to simulate the transport of Asian rust urediniospores to the state of Paraná, by cold fronts. Several studies in the literature perform this approach to simulate the transport of physical properties in fluid media (Romeiro et al., 2011; Pardo et al., 2012; Leelossy et al., 2014; Saita et al., 2017; Romeiro et al., 2018). Thus, the present study developed numerical simulation to predict the atmospheric transport of *P. pachyrhizi* urediniospores from places that grow soybean in the off-season (i.e., Paraguay) to Paraná state, Brazil.

## 2. Materials and methods

2.1 *About the state of Paraná*

The state of Paraná is located in the south of Brazil, bordered on the north with São Paulo, on the south with Santa Catarina, on the southeast with the Atlantic Ocean, on the southwest with Argentina, on the west with Paraguay and on the northwest with Mato Grosso do Sul, (Figure 1). The territorial area of the state is approximately 119,554 km$^2$, which represents 2.34% of the Brazilian territory (COPEL, 2007).

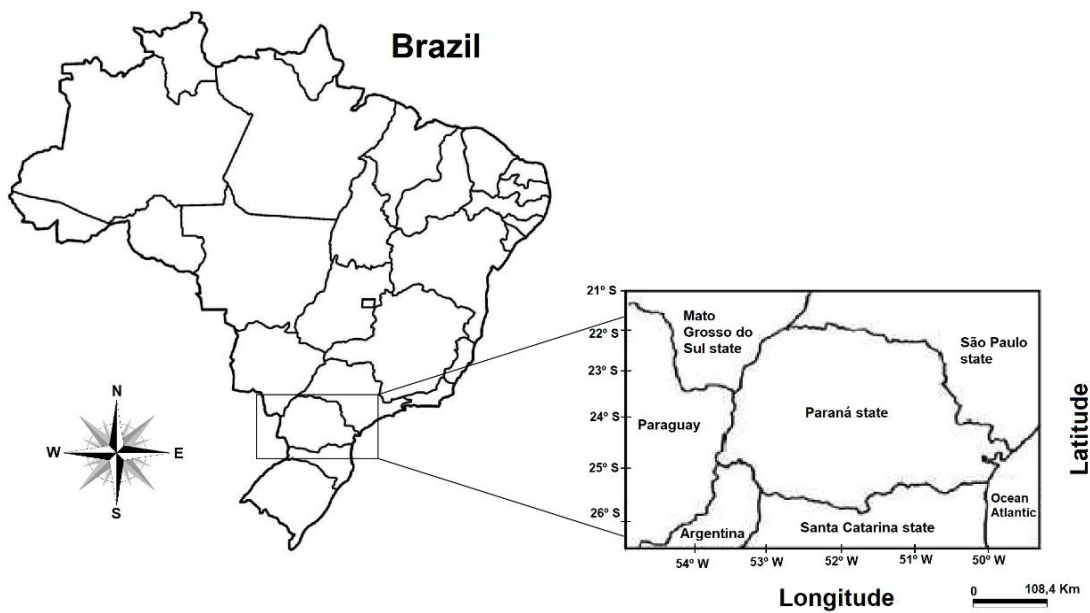

**Figure 1** Geographic location of the state of Paraná in the Brazilian territory. The highlighted square region is the domain of numerical simulations. **Source:** The authors.

2.2 *Mathematical model*

The mathematical model adopted in this study that describe the bidimensional atmospheric transport of *P. pachyrhizi* urediniospores in Paraná state is composed by a partial differential equation that considers diffusive, convective and reactive effects (Martin and McCutcheon, 1998; Romeiro et al., 2011; Pardo et al., 2012; Leelössy et al., 2014; Romeiro et al., 2018; Saita et al., 2022), as given below:

$$\underbrace{\frac{\partial C}{\partial t}}_{\text{temporal term}} = \underbrace{D\nabla^2 C}_{\text{diffusive term}} - \underbrace{\vec{\nabla} \cdot (C\vec{v})}_{\text{convective term}} - \underbrace{\lambda C}_{\text{reactive term}} \qquad (1)$$

where $D = 10^{-2}$ km²/h denotes the diffusion coefficient of urediniospores suspended in the air and $\lambda = 0.12$ h$^{-1}$ is the mortality rate of urediniospores during atmospheric transport. The quantities $D$ and $\lambda$ are constant (see Table 1).

Note that Equation (1), when $\lambda = 0$, corresponds to the classical convective diffusion wave equation. The deduction of the convective diffusion wave equation is obtained from the mass conservation principle for the concentration of the transported quantity $C(x, y, t)$, as shown in (Socolofsky and Gerhard, 2005). Note that the total velocity of the transported quantity may be split into convective velocity and into diffusive velocity. The first term on the right side of (1) describes the diffusive transport of urediniospores, modeled by means of Fick's Law (Martin and McCutcheon, 1998). The second term on the right side of (1) describes the convective transport of urediniospores due to cold fronts. Note that during cold fronts, convective transport is dominant. The last term on the right side of Equation (1) corresponds to an exponential decay term for the concentration of urediniospores, which models their mortality rate during atmospheric transport.

The function $C(x, y, t)$ represents the concentration of urediniospores and depends on two spatial variables $x$ and $y$, and a time variable $t$. In the convective term, the quantity $\vec{v} = v_x(x,y,t)\vec{i} + v_y(x,y,t)\vec{j}$ is the speed field of cold fronts. Finally, Nabla and Laplacian differential operators are given respectively by: $\vec{\nabla} = \left(\frac{\partial}{\partial x}\right)\vec{i} + \left(\frac{\partial}{\partial y}\right)\vec{j}$ and $\nabla^2 = \vec{\nabla}\cdot\vec{\nabla} = \frac{\partial^2}{\partial x^2} + \frac{\partial^2}{\partial y^2}$.

2.3 *Initial condition*

The initial condition of urediniospores concentration was estimated with the areas with winter soy presence in Paraguay. Soybean cultivation in Paraguay is restricted to the border region with Brazil, the western region of Paraguay. In this article, the highlighted square region in Figure 1 is considered as the domain of numerical simulations. Note in Figure 1 the western region of Paraguay considered for the initial urediniospores data. In Paraguay, until 2019, there were no restrictions for growing soybean in winter.

As the aim of this article is to verify whether Asian rust contamination in Brazil is introduced by cold fronts which bring urediniospores from Paraguay, residual pathogens are not considered in any area of Brazil, including the state of Paraná. It is considered that in the entire Brazilian region of numerical simulation (see the highlighted square region in Figure 1) the concentration of urediniospores is null. On the other hand, the western region of Paraguay considered in the numerical simulations, where there is soybean planting and there is no regulatory measure that prohibits the cultivation of soybeans (soybean-free period), in this region, it is considered that the urediniospore concentration is proportional to the soy planted density. The greatest number of soy plantations in Paraguay occur in the border region with Brazil. Moving away from the border, the density of soy plantations in Paraguay decreases. The data used are from the

Mesa de la Roya website (http://www.mesadelaroya.com/index.php/blog/79-cultivos-trampas-para-deteccion-de-roya-de-soja).

2.4 *Speed field*

The standard wind regime that occurs in the state of Paraná is characterized by trade winds (COPEL, 2007). However, this wind regime is also influenced by the passage of cold fronts that intensify in winter (June - September) and spring (September - December). Cold fronts have more intense winds due to hot air masses, coming from the North or Northeast of Brazil, colliding with polar air masses. Figure 2 schematically represents a possible velocity field due to a cold front (COPEL, 2007).

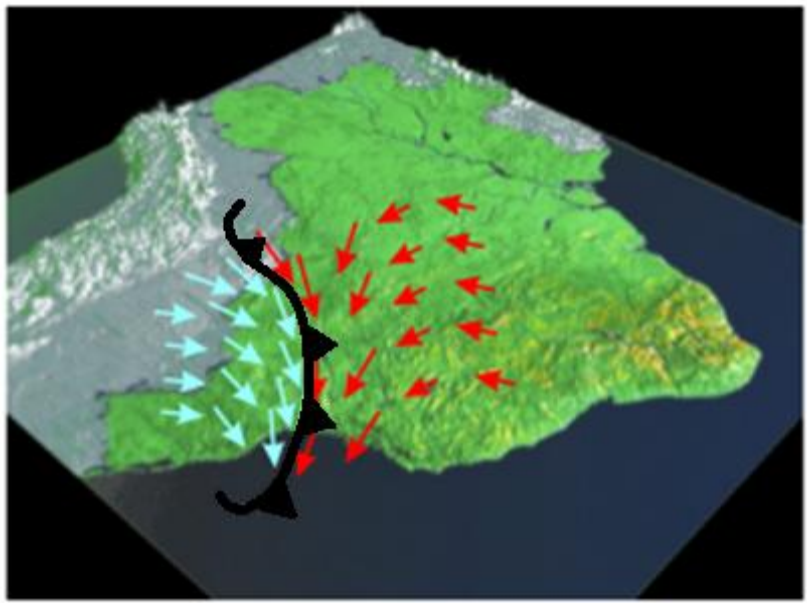

**Figure 2** Schematic representation of a possible velocity field due to a cold front passing through the state of Paraná. **Source:** Copel (2007).

For the simulations performed in this work, the speed field of the convective term, Equation (1), is given empirically by the characteristics of air speed and direction in a cold front. The parameters were obtained from the National Space Research Institute (INPE) available at http://tempo.cptec.inpe.br/boletimtecnico/pt. The characteristics of velocity and direction of cold fronts occurred between October 2018 and February 2019 were considered.

2.5 *Urediniospores mortality*

The mortality rate of urediniospores is unknown during their transport by cold fronts. Factors such as temperature, relative humidity and solar radiance influence urediniospores survival. *P. pachyrhizi* and *P. triticina* urediniospores exposed to sunlight drastically reduced germination and germ tube elongation after 2.5 h (Buck et al., 2010; Young et al., 2012). However, in this work, it was considered that during the cold front the effects of UV radiation on the mortality of urediniospores are null. Note that during transport by cold fronts, urediniospores are shielded from UV radiation by thick cumulus clouds. In this context, the moisture and temperature were used as parameters. In laboratory conditions, urediniospores begin to germinate three hours after the inoculation in agar-water, with the maximum germination occurring 6 to 7 hours after (Blum et al., 2015). It was assumed that urediniospores that germinate before the contact with susceptible host are non-viable for infection (dead).

In our mathematical modeling, a constant mortality rate for urediniospores is assumed. Based on laboratory measurements at the State University of Londrina, the constant mortality rate that best fits the experimental data is $\lambda = 0.12$ h$^{-1}$ (data are unpublished). Furthermore, this value for the mortality rate for urediniospores is also consistent with the results obtained in (Blum et al., 2015).

Due to the soybean free period, soybean planting is allowed after September 15th in Paraná state. The soybean sowing period was considered to provide information about the host susceptibility to infection. For the simulations, it was assumed that for the period until 30 days after sowing the probability of *P. pachyrhizi* infection is null. The assumption was based on the study of (Schmitz and Grant, 2009) that showed that wetness in the first soybean stages reduced the probability of infection. Historically, the first detected cases of Asian rust in Parana state occurred in early November, more than 30 days after sowing, according to the Anti-rust Consortium (available at http://www.consorcioantiferrugem.net//main).

2.6 *Simulations*

The mathematical model (1) was discretized by the finite difference method. The implicit linear algebraic system resulting from discretization was solved by the Gauss-Seidel iterative method (Romeiro et al., 2011; Pardo et al., 2012; Saita et al., 2017; Romeiro et al., 2018). The simulations were performed in GNU Octave, version (4.2.2). The time considered in the simulations is three hours, which represents an average of the local duration of the cold fronts that were collected in this work (INPE lab, SIMEPAR lab). For each cold front, the initial condition of the concentration of urediniospores was given from the transport that occurred on the last cold front. Parameters used for numerical simulations are presented in Table 1.

**Table 1** Values of the parameters adopted in the diffusive, convective and reactive terms of the model that simulates the transport of urediniospores in the state of Paraná.

| Parameter | Description | Value | Unit | Reference |
|---|---|---|---|---|
| $C$ | Concentration | 0 to 1 | Dimensionless | |
| $T$ | Average duration of the cold fronts | 3 | h | INPE lab SIMEPAR lab |
| $\lambda$ | Mortality rate | 0.12 | $h^{-1}$ | UEL lab |
| $D$ | Diffusion coefficient | $10^{-2}$ | $km^2\,h^{-1}$ | Tsuda et al. (2013) |
| $\vec{v}(x, y, t)$ | Convective speed field | 40 – 90 | $km\,h^{-1}$ | INPE lab SIMEPAR lab |

2.7 *Validation*

The results of simulations were validated according to the daily number of cases reported by the Anti-rust Consortium. The Anti-rust Consortium (available at http://www.consorcioantiferrugem.net//main) is a public-private collaboration programme which monitors soybean rust occurrence in Brazil. The networking collaboration is composed of certificated laboratories and researchers who upload occurrence cases in the system after the investigations of the samples (Godoy et al., 2016).

**3. Results**

During the studied period, direction and speed characteristics of nine cold fronts that intercept the state of Paraná were considered. Table 2 presents the main characteristics of these cold fronts.

**Table 2** Information from the nine cold fronts observed in the 2018-19 season.

|  | Occurrence date | Predominant wind direction | Convective velocity field figure |
|---|---|---|---|
| Cold front 1 | 25/10/2018 | west to northeast | Figure 3a |
| Cold front 2 | 27/10/2018 | southwest to southeast | Figure 3b |
| Cold front 3 | 01/11/2018 | west to northeast | Figure 3c |
| Cold front 4 | 23/11/2018 | southwest to southeast | Figure 3d |
| Cold front 5 | 01/12/2018 | northwest to northeast | Figure 3e |
| Cold front 6 | 22/12/2018 | west to east | Figure 3f |
| Cold front 7 | 05/01/2019 | west to east | Figure 3g |
| Cold front 8 | 15/01/2019 | west to northeast | Figure 3h |
| Cold front 9 | 01/02/2019 | west to southeast | Figure 3i |

Figures 3a-3i present the velocity fields (directions and intensities) of the nine cold fronts considered in this study (INPE lab). The maximum wind speed registered on cold fronts was 90 km/h, as shown in the color scale of Figure 3.

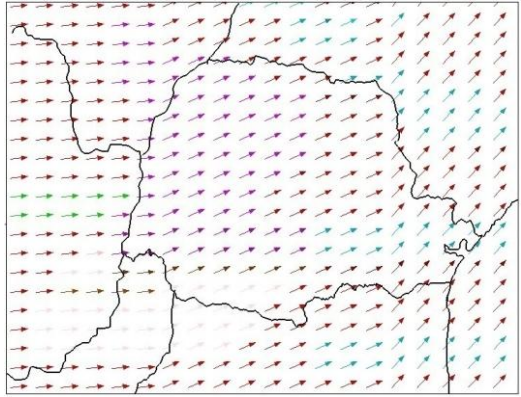

(a)

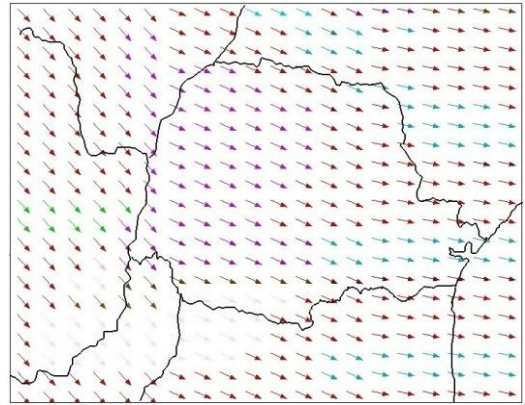

(b)

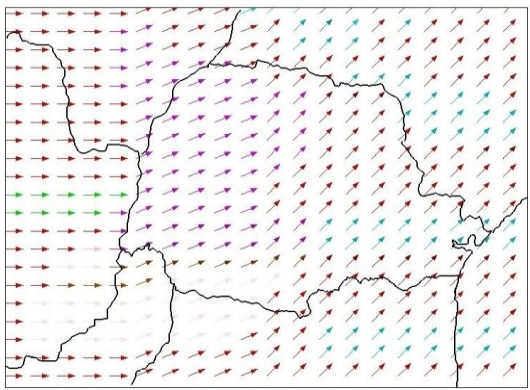

(c)

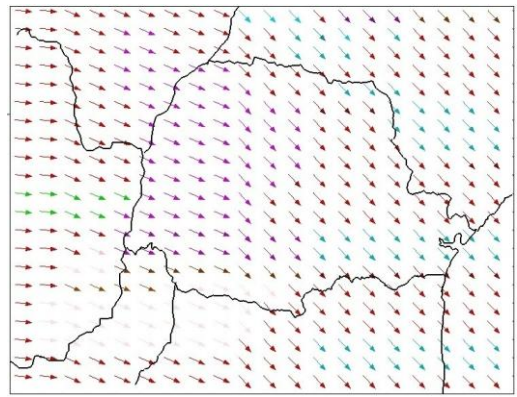

(d)

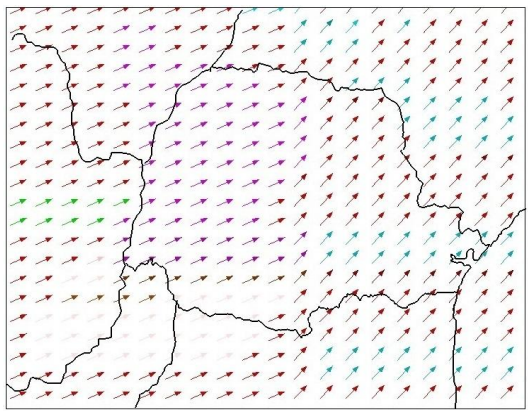

(e)

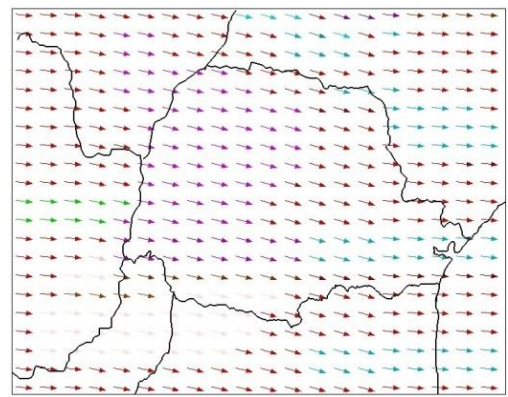

(f)

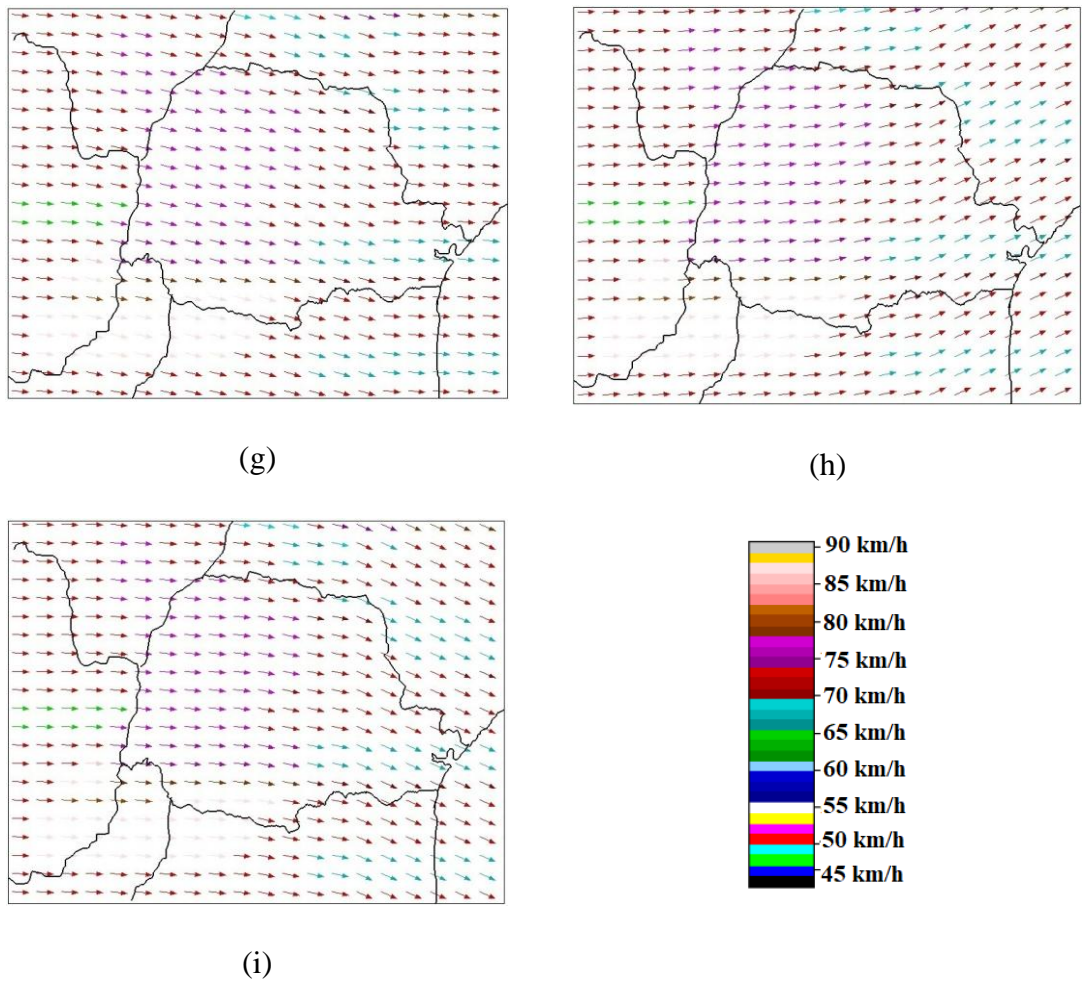

(g) (h)

(i)

**Figure 3** Wind direction and velocity of cold fronts that intercepted Parana state in October 25th and 27th, (a) and (b) figures; November 1st and 23th, (c) and (d) figures; December 1st and 22th, (e) and (f) figures; January 5th and 15th, (g) and (h) figures; and February 1, (i) figure. The data were provided by INPE lab. **Source:** Adapted from INPE lab.

The mathematical model, Equation (1), describes the variation of concentration of urediniospores $C(x,y,t)$ in the highlighted square region in Figure 1, at position $(x,y)$ and at time $t$, given initial and boundary conditions. In this context, Equation (1) is a bidimensional atmospheric transport differential equation of *P. pachyrhizi* urediniospores. The velocity fields of Figures 3(a)-(i) are input in Equation (1), in the convective term. Thus, the numerical simulations of Equation (1) describe the transport by diffusive (molecular diffusion) and convective (cold fronts) effects of *P. pachyrhizi* urediniospores from Paraguay to Paraná.

Figure 4 shows the urediniospores being transported by the cold front October 25[th] (Figure 3a), considering time periods of 15 minutes. The colour gradient shows that the regions in shades of red are those with the highest concentration of urediniospores, while the regions in shades of blue represent those with the lowest concentrations.

Explaining the simulations presented in Figure 4. The government of Paraguay does not release quantitative data on the contamination of Paraguayan soybeans by *P. pachyrhizi* urediniospores. In this context, it is considered that the urediniospores concentration $C(x,y,t)$ in Paraguay is proportional to the density of soy plants, ranging from 0 to 1, as described in Table 1. In Figure 4, frame $t = 0$ describes the initial condition of the numerical simulation. Note that in frame $t = 0$, the highlighted square region in Figure 1 has a non-zero concentration of urediniospores only where soybean plantations occur in Paraguay. In the other frames, Equation 1 describes the transport of the concentration of urediniospores $C(x,y,t)$ in the studied region later.

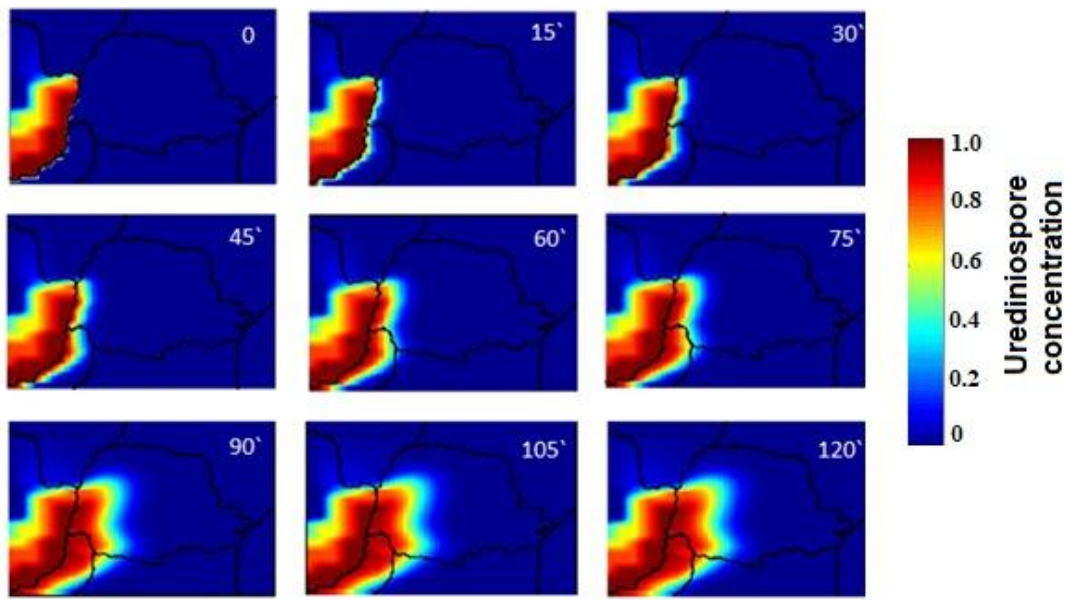

**Figure 4** Numerical simulation of concentration of urediniospores transported by cold front October 25[th]. The figures represent frames with time step of 15 minutes. **Source:** The authors.

According to the Anti-rust Consortium, soybean sowing in this region occurs on average between September 15$^{th}$ and October 1$^{st}$ (Johann et al., 2016). Figure 5b shows the final dispersion of urediniospores due to the cold front on October 25$^{th}$. It also shows (black dots) the places where the urediniospores had already been observed to date. The pink dots are the locations of the main cities of Paraná. Observe the consistency between the observed data (black dots) and the numerical simulation.

The second cold front (Figure 3b) occurred on October 27$^{th}$. Numerical results indicated that the urediniospores were transported to the region south and southeast of state of Paraná (Figure 5c). In November, the first cold front occurred in 1$^{st}$ with direction from southwest to northeast (Figure 3c). The numerical simulation presents transport of urediniospores to the east and north of the state (Figure 5d). The concentration of urediniospores resulting from the spread that occurred on the cold fronts of October 25$^{th}$ and 27$^{th}$ and November 1$^{st}$ reached regions that showed early soybean sowing and that present susceptible hosts. In the same period, there was the first report of the occurrence of the disease on the Anti-rust Consortium Portal, which occurred on November 5$^{th}$ in the city of Ubiratã, Paraná. About three days later, the disease was also reported in the cities of Sao Miguel do Iguacu (November 8$^{th}$), Nova Cantu (November 8$^{th}$), Campo Mourão (November 8$^{th}$) and Juranda (November 9$^{th}$).

Figure 5 shows the numerical simulations of the atmospheric transport of urediniospores in the state of Paraná, when the occurrence of the nine cold fronts is considered. The pink dots indicate the location of the largest cities in Paraná such as Foz do Iguaçu, Cascavel, Maringá, Londrina, Ponta Grossa and Curitiba. Black dots indicate the cases of soybean rust confirmed by the Anti-rust Consortium in the period, until the occurrence of the cold front studied.

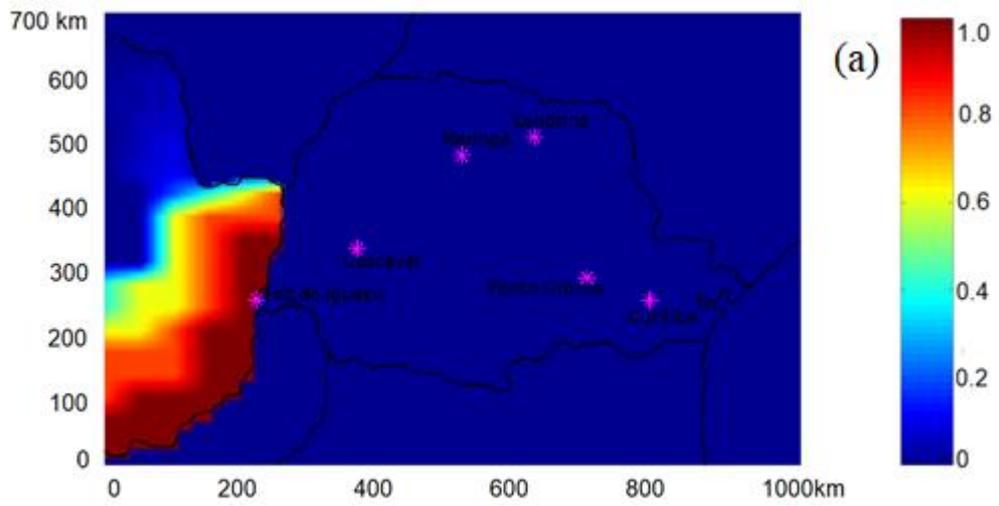

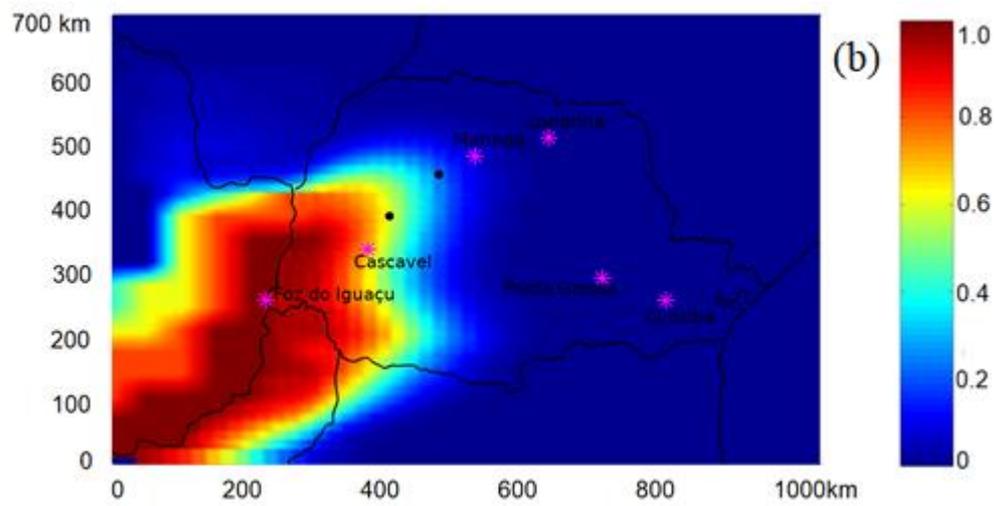

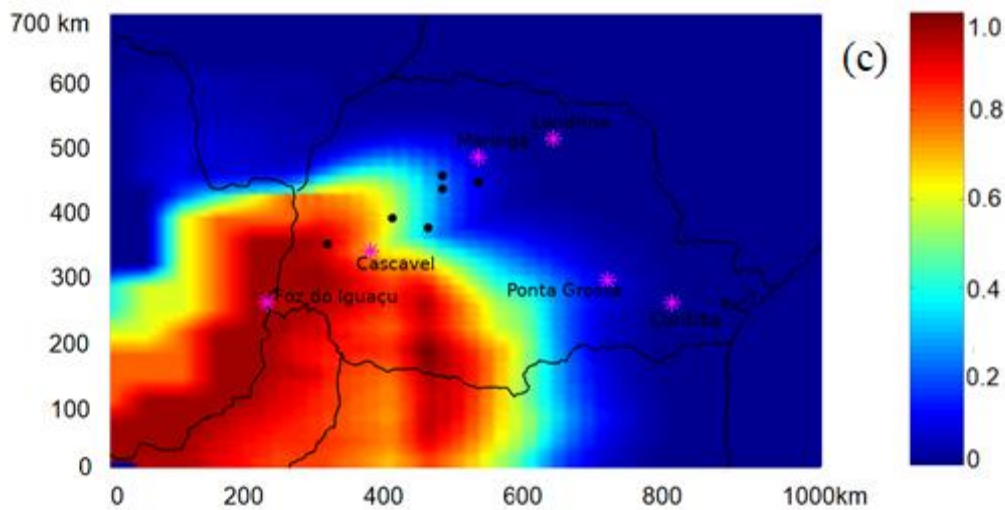

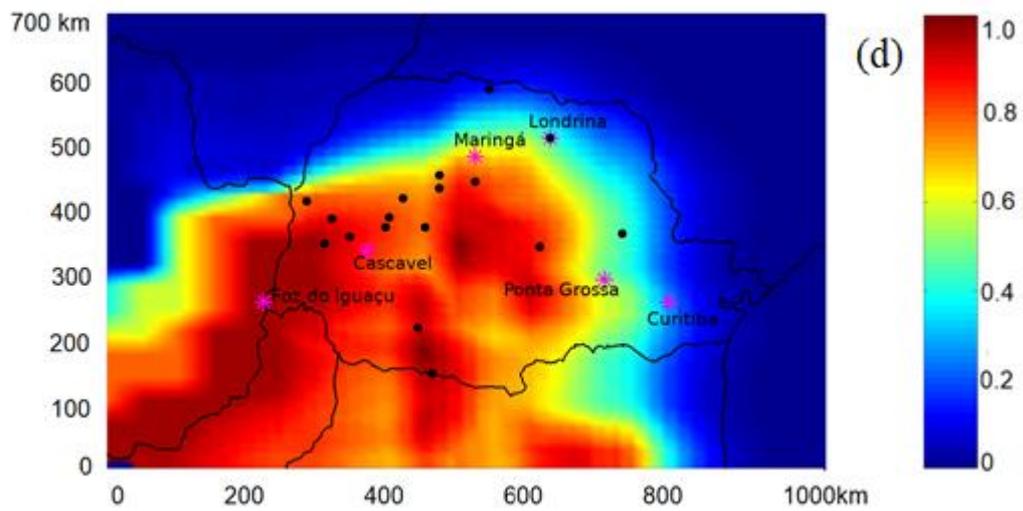

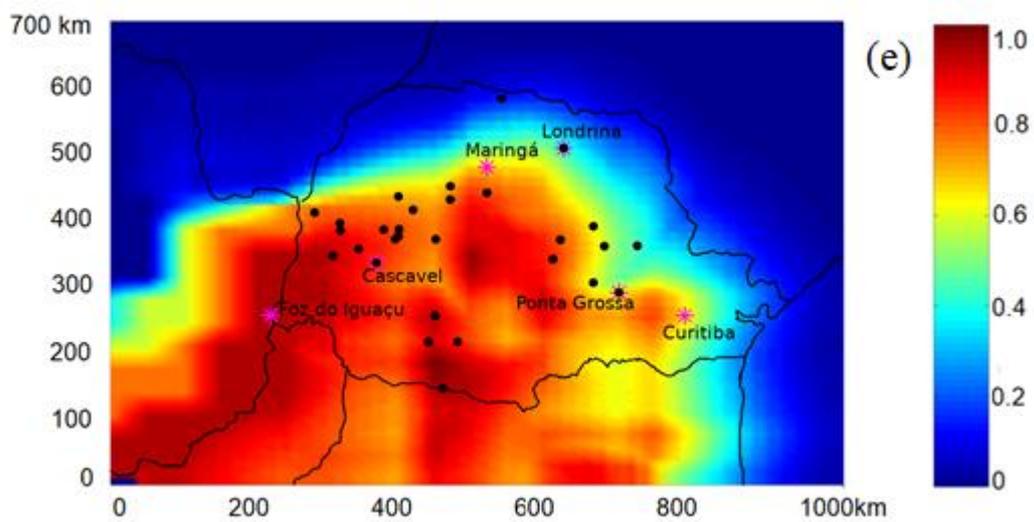

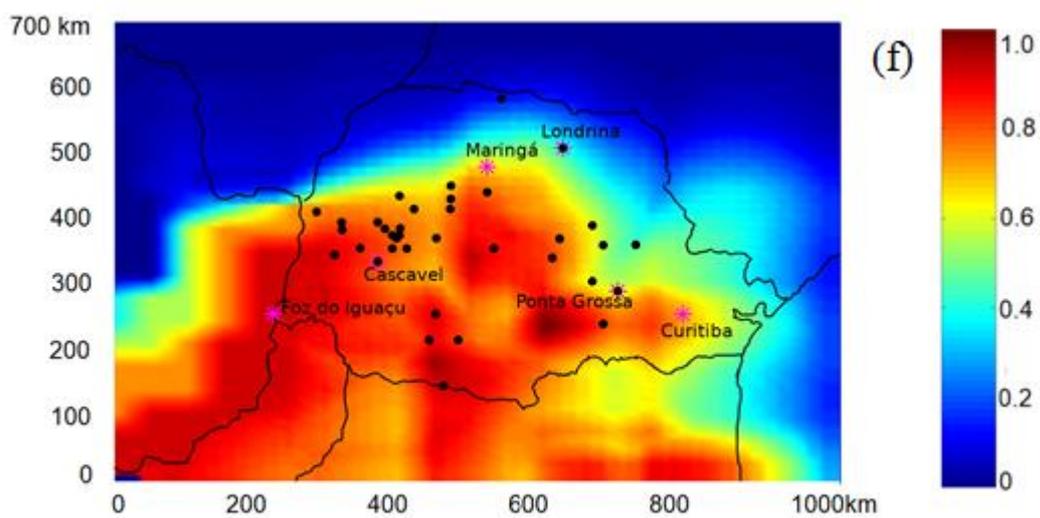

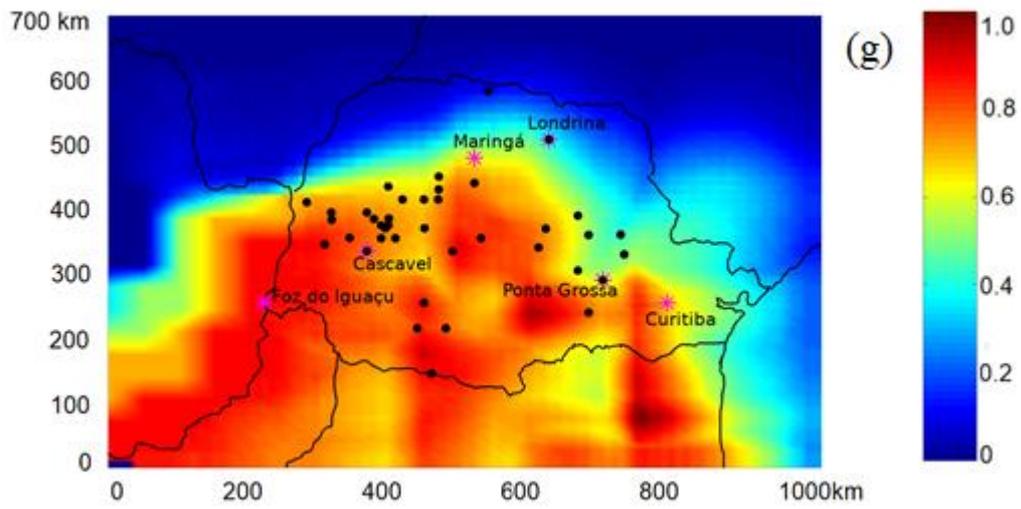

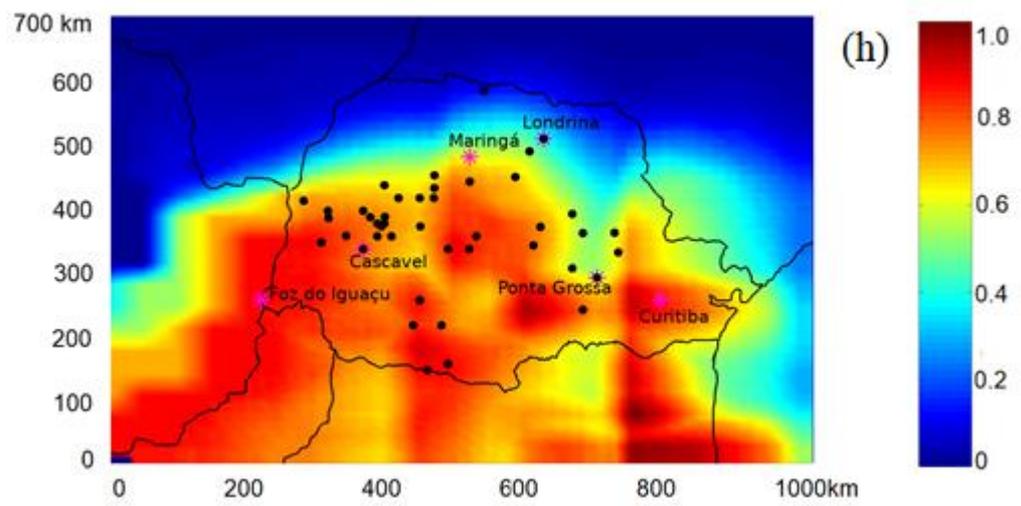

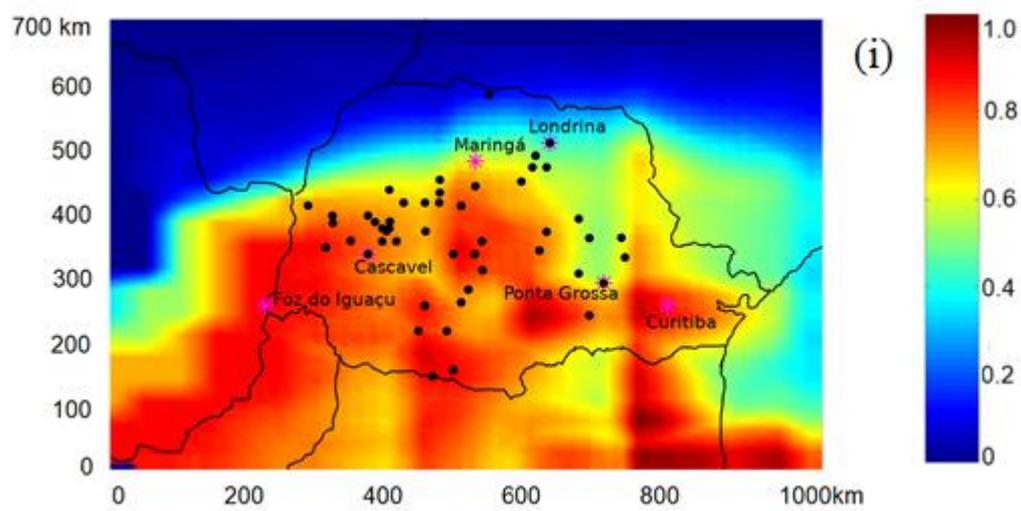

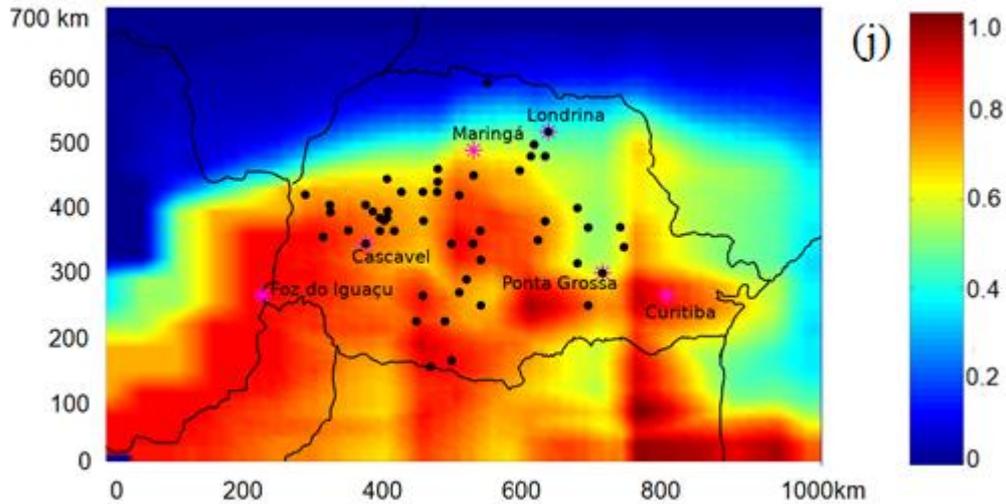

**Figure 5** Numerical simulation of the temporal and spatial concentration of urediniospores in the state of Paraná due to cold fronts occurred across the region between October 2018 and February 2019. Figure 5a shows the contaminated regions in Paraguay. Figure 5b shows the transport of urediniospores due to the cold front on October 25, 2018. Figure 5c on October 27, 2018. Figure 5d on November 1, 2018. Figure 5e on November 23, 2018. Figure 5f on December 1, 2018. Figure 5g on December 22, 2018. Figure 5h on January 5, 2019. Figure 5i on January 15, 2019 Figure 5j on February 1, 2019. The black dots show the temporal and spatial evolution of the infection observed by the Anti-rust Consortium in the state of Paraná. The pink dots are the location of the largest cities of Paraná. **Source:** The authors.

The experimental observations (black dots in Figures 5a-5j) corroborate the results obtained in the numerical simulations of the mathematical model, considering that the visual detection of symptoms of Asian rust occurs after the latency period that can vary from 7 to 15 days (Danelli and Reis, 2016). Braga et al. (2020) presented a latent period for samples collected in these regions of Paraná between 10 and 15 days.

The results presented in Figure 5, showed that consistent with the physics of the problem under study, the urediniospores were transported following the same direction of the airflow of cold fronts. In the simulations, the concentration of urediniospores transported by the first cold front (25/10/2018) is an initial condition for the second cold front (27/10/2018), and so on. The initial conditions were established in this way, so the regions that were contaminated are taken into account when calculating the dispersion of the urediniospores for the next cold front. Finally, in the simulations, no other effects on the dispersion of urediniospores were considered, such as fungicide applications in the infected regions.

**4. Discussion**

The results presented by simulation suggest that urediniospores of *P. pachyrhizi* can be transported from soybean areas of Paraguay to Paraná state in Brazil by cold fronts.

The dispersion of the pathogen is the central key of development of an epidemic disease that are connected with cycle of infection, multiplication of inoculum and dispersal of inoculum to new infection sites (Aylor, 1990; Pivonia and Yang, 2004). The dissemination for long-distance depends on factors about the pathogen (survival of spores during exposure to temperature, humidity, and UVB radiation), host (presence of susceptible host) and environment (atmospheric turbulence, stability, and wind speed). According to Aylor (2003), the practical limit for long-distance dispersion of a plant pathogen depends strongly on its fecundity and ability to survive in the atmosphere. Levetin (2007) cited that spores that are adapted to airborne dispersion are often much more resistant to environmental stress than are the parent hyphae.

In this context, the concentration of the initial content of urediniospores represents a key factor for the success of the dispersion. Without any soybean growth restriction in Paraguay, the number of rust cases in Paraná, season 2018/19, until December 31$^{th}$ was 48 (Anti-rust Consortium). After the soybean-free period implementation in Paraguay in the winter of 2019, the first case of soybean rust was reported on December 5$^{th}$ and the total number of rust cases in Paraná until December 31$^{th}$ was 13, the lowest since the creation of Anti-rust Consortium. The evidence suggests the reduction in initial inoculum. The increase in the number soybean rust cases after December indicates the short-distance movement.

Otherwise, the reduction in the number of cases can be explained by other factors than concentration of the initial content of inoculum, such as the environmental conditions for the disease occurrence due to the El Niño/La Niña Southern Oscillation (ENSO) phenomenon (Minchio et al., 2016, 2018; Nóia Júnior et al., 2020), sowing time, and resistance of cultivars that give unfavorable conditions to the fungal infection (Braga et al., 2020, Xavier et al., 2017). In addition, questions persist about the initial concentration of spores that are dispersed, and about which fraction of spores produced by any lesion that will participate in short and long-distance dispersion. These questions require further study.

According to Golan and Pringle (2017) study and development of novel approaches to evaluate the spore trajectories and survival probability are "critically needed". The validation of the mathematical model with soybean rust cases by the Anti-rust Consortium can be one of the alternatives for simulation models. In this context, through numerical simulations, it is intended to show evidence that the cases of Asian rust found in the state of Paraná, during the 2018/19 season, were caused by urediniospores from Paraguay, transported by cold fronts. The results presented can help farmers in making decisions about fungicide applications. In addition, integrated strategies have been suggested to assist in the management of the disease, such as early sowing (Dias et al., 2014; Koga et al., 2014), resistant cultivars, field monitoring and others (Fantin et al., 2019).

## 5. Conclusion

In this work, a partial differential equation with diffusive, convective and reactive terms was used to simulate the two-dimensional transport of Asian soybean rust spores in the state of Paraná. It is assumed that the spores come from soy producing regions of Paraguay, and that the transport of these spores is carried out through cold fronts.

The numerical simulations carried out show that there is a spatial and temporal agreement with the data measured by the Anti-rust Consortium on the cases of Asian rust in the state of Paraná, from October 2018 to February 2019. They also support a discussion about an effective implementation of a soybean-free period on a continental scale.

## Acknowledgments

This study was financed in part by the Coordenação de Aperfeiçoamento de Pessoal de Nível Superior – BR (CAPES) - Finance Code 001.

**Compliance with ethical standards**

**Conflict of interest:** The authors declare that they have no conflict of interest.

**Human participants and/or animals:** This article does not contain any studies with human participants or animals performed by any of the authors.

**Informed consent:** Informed consent was obtained from all individual participants included in the study.